\documentstyle[aps]{revtex}
\begin{document}
\begin{title}
{\bf Coulomb Drag Between Quasiballistic Quantum Wires:\\
an Indication of Non-Fermi-Liquid Behavior.}
\end{title}
\author{  P. Debray, V. Zverev\cite{vol} ,
O. Raichev\cite{ol} , R. Klesse$^{\star}$,
P. Vasilopoulos$^{\diamond}$,
and M. Rahman$^{\ast}$\\
\ \\}

\address{Service de Physique de l' \'{E}tat Condens\'{e}, CEA Saclay,
91191 Gif-sur-Yvette, France\\
$^{\star}$Universit\"at zu K\"oln, Institut f\"ur Theoretiche Physik,
D-50937 K\"oln, Germany\\
$^{\diamond}$Concordia University, Department of Physics,
Montr\'{e}al, Qu\'{e}bec, Canada, H3G 1M8 \\
$^{\ast}$Department of Physics, University of Glasgow, Glasgow, G12 8QQ, UK}


\maketitle

\begin{abstract}

The Coulomb drag between two spatially separated, 2 $\mu$m long
lithographically defined quantum wires has been studied
experimentally in the absence of interwire tunneling.
The drag resistance $R_D$ shows peaks when the 1D subband
bottoms of the wires are aligned and the Fermi wave vector $k_F$ is small.
$R_D$ decreases exponentially with  the interwire separation $d$.
In the temperature range $0.2K \leq T\leq 1K$ the drag signal shows the
{\it power-law dependence} $R_D\propto T^x$  with $x$ ranging from -0.61
to -0.77 depending on the magnitude of $k_F$. We interpret our experimental
results in the framework of the Tomonaga-Luttinger liquid theory.\\


\end{abstract}


One-dimensional (1D) electron systems have recently been the focus of
considerable
attention since they are expected to show unique transport
properties associated with the Coulomb interaction between carriers.
In 1D systems this interaction
modifies  the ground state and the elementary excitations
considerably and the systems
are theoretically described in terms of a Tomonaga-Luttinger (TL) liquid
rather than a Fermi-liquid (FL) model [1]. Recently attempts have been
made to test some of the predictions of the TL theory in
1D systems, such as lithographically defined quantum wires, which
proved to be convenient for transport measurements and allow the variation
of the relevant parameters 
in a wide range. Though it has been argued [2]
that according to the TL theory the single-mode conductance of a
ballistic quantum
wire should deviate from its fundamental value of $G_0=2e^2/h$, the majority
of the wires investigated do not show such a deviation, and where it is present
[3], it still lacks a consistent explanation in terms of a TL liquid.
The reasons that make the measured conductance independent of the interaction
have been discussed by many authors [4], who emphasized the role of the
FL reservoirs to which the wires are connected and of the screening
effects. Therefore, one should search for other, more ingenious ways
to obtain experimental evidence of a TL liquid behavior. One such way would  be
to use the Coulomb drag (CD) between parallel quantum wires in line with its
recent TL liquid descriptions [5-7] which strongly suggest that it can
be used to probe this TL liquid behavior experimentally.

Experimental and theoretical studies of the CD between two-dimensional (2D)
electron layers, recently reviewed [8], established that the drag resistance $R_D=-V_D/I$, where
$V_D$ is the drag voltage developed in the drag layer as a response
to the current $I$ flowing through only the drive layer, decreases with
the decreasing temperature $T$. This behavior is consistent with a FL
description of the electron system, since the restrictions imposed by
the momentum and energy conservation laws suppress the
probability of electron-electron scattering at smaller $T$.
Similarly, within a FL description of the drag between 1D systems
\cite{9}-\cite{10a}
such restrictions have even stronger consequences: the low-temperature
drag response is maximal when the energy levels of the wires are aligned so
that the Fermi velocities are nearly equal. If the alignment is perfect,
the drag linearly decreases with temperature,
otherwise the decrease is exponential.

In contrast with this behavior of the CD in 2D or 1D
systems, its temperature dependence
changes completely when the electron subsystems in the two
wires behave like TL liquids and the interwire momentum transfer is
strongly modified by electron-electron interactions. Roughly speaking,
the drag results from backscattering of density excitations in one wire
from density fluctuations in the other one. Therefore, the present
situation bears some similarity [7] to a TL liquid with backscattering
by an impurity. Since for a repulsive interaction the effective backscattering
strength of the impurity increases with decreasing temperature and eventually
diverges at $T=0$ [12], one should expect that similarly the drag becomes
enhanced at low temperatures until, at $T=0$, interlocked charge density waves
form and the drag-resistance diverges too.
(This divergence or {\em absolute drag} [6] applies only to
wires that are infinitely long.
For finite wires, at sufficient low temperatures the drag becomes suppressed
due to the influence of the contact reservoirs [13].) Though this
strong-coupling regime may remain elusive experimentally, the increase of the
drag with decreasing $T$ 
in a characteristic power-law fashion [7]
may serve as a signature of the TL behavior.

Recently, we have reported [14] experimental evidence of the CD between 1D
electron systems. In the present work we report comprehensive studies of this
effect, successfully explain the observed features in the temperature
dependence of $R_D$ in the framework of the TL  theory, and show
how these features 
can be used to probe a TL liquid. The details are as follows.

{\it Measurements}.
The quantum-wire devices were fabricated from a 2D electron layer
(with electron mobility $\mu=$100 m$^2$/V$\cdot$s and electron density
$n=2.7 \times 10^{15}$ m$^{-2}$ at 4.2 K)
located at the interface of a AlGaAs/GaAs heterostructure 80 nm below the
surface.
The lithographically fabricated planar structure consists of three
independent surface Schottky gates T, M, and B (Fig. 1), with 50 nm wide
middle gate and 0.25 $\mu$m wide slits of length $L=2$ $\mu$m.
By appropriate voltage biasing of these gates, it was possible to vary
the widths of the wires and their separation. Electrical characterization
of the wires was carried out by measuring their conductance at 60 mK. The
conductance was measured of one wire at a time, and of both of them
simultaneously, as a function of wire width to establish
the ballistic nature of electron transport and to check
if both wires had identical transport behavior. The two wires were
found to have nearly identical conductance staircases with a small
difference in the pinch-off voltages which could be compensated for
by introducing an appropriate voltage shift between the gates T and B.
The observed conductance showed characteristic features of
{\it ballistic transport}. However, the conductance staircases were not
sufficiently well defined due, very likely, to deviations from adiabaticity
at the constriction openings and scattering in the wires caused by gate
edge roughnesses. The application of a magnetic field $B < 1$ T perpendicular
to the plane of the device improved the adiabaticity and suppressed
the scattering, producing fairly well-defined plateaus (Fig. 2).
This, together with the fact that we did not observe any sharp peaks in
the pinch-off regime and/or resonant oscillations on the plateaus,
confirmed that we were dealing with ballistic transport in quantum wires
free of embedded impurities or dots. The information
obtained from the above characterization made it possible to choose,
as required, the location of the Fermi level $E_F$ in a specific 1D subband
as well as the relative alignment of the 1D subbands belonging to the
two wires. For measurements of the CD effect, the upper wire
was chosen as the drive wire and the lower one was the drag wire (Fig. 1).
The gates M and B were appropriately biased with
voltages $V_M$ and $V_B$, respectively, to have $E_F$ slightly above the
bottom of the lowest 1D subband of the drag wire. The influence of the
voltage $V_T$ applied to the upper gate on the width of the drag wire
was found to be insignificant.
A driving voltage $V_{DS}$, low enough to be within the linear regime of
transport, was applied to the drive wire to send a current $I$ through it.
No current was allowed to flow in the drag wire. $I$ and the drag
voltage $V_D$, opposite in sign to $V_{DS}$, were measured simultaneously
as $V_T$ was swept.
To ensure the absence of tunneling during the measurements, the
tunneling current across the middle gate, between the drain of the drive
wire and the source of the drag wire, was measured as a function of
$V_M$. The tunneling could be neglected for $V_M < -0.7$ V; accordingly
$V_M$ was chosen to be less than $-0.7$ V,
except when studying the effect of the interwire separation on the drag.
Contrary to Ref. [14], the application of a magnetic field was not
necessary to suppress the tunneling because of the much better quality
of the Schottky gates of the devices used in this work.
The above procedure
was used to measure $V_D$ at different temperatures and in magnetic
fields applied perpendicular to the plane of the device.

{\it Results and discussion}.  In Fig. 2 we present the drag voltage $V_D$ as a function of the
upper gate voltage $V_T$ for zero magnetic field ($B=0$) and fixed values
$V_M$, $V_B$, and $V_{DS}$ given in the caption. We also present the
drive wire conductance $G$ for $B=0$ or finite $B$ as indicated in
the caption. As can be seen, $V_D$ shows two prominent peaks,
one at $V_T=-1.21$ V and one at $V_T=-1.10$ V; they occur in the
rising parts (steps) between the conductance plateaus. We have also
found that $V_D$ is a linear function of
$V_{DS}$ up to $V_{DS} \simeq $400 $\mu$V, beyond which $V_D$ starts
to show sublinear behavior, possibly caused by a change of
subband population in the drive wire [15]. The inset demonstrates
this effect for the first peak voltage $V_D=V_D^{max}$. Measurements in a
field $B=0.86$ T show identical behavior except that the magnitude of
$V_D$ is enhanced almost by a factor of 3, which we attribute to
the enhancement of the density of 1D states (at  this  field
the magnetic length is comparable to the wire width).
The results shown are representative of typical
device behavior. In this work we focus on 1D transport in
the fundamental mode and discuss mainly the region of the first peak.

The origin of the first peak in $V_D$  can be understood by taking into
account that the CD is proportional to the Coulomb-assisted backscattering
probability in 1D systems. This probability is enhanced when the 1D
levels are aligned \cite{9,10} and the Fermi wave vectors $k_F$ in the wires
are small so that interwire momentum transfer $\hbar q \simeq 2 \hbar k_F$
is small too. As seen from Fig. 2, this peak corresponds
to these conditions. (In a similar way one can show
that the second peak in $V_D$  occurs
when two different subband bottoms from the two wires line up \cite{10a},
\cite{13}.)
The backscattering probability is proportional to $[K_0(2k_F d)]^2$,
where $K_0$ is the modified Bessel function and $d$ the interwire distance
\cite{10}, \cite{10a}. To
check the reliability of this expression and to estimate the Fermi wave vector
corresponding to the peak value of $V_D$ we measured the
dependence of the drag signal on $d$ by changing the middle gate
voltage $V_M$, which moves the depletion region in a nearly linear
way and thus changes $d$. In Fig. 3 we show the peak value of
$V_D$ and $R_D$ as a function of $V_M$. To carry out these measurements, each
time $V_M$ was changed, $V_B$ was adjusted to maintain the same width of the
drag wire so that $E_F$ was always just above the bottom of the
lowest 1D subband.
The dependence of $R_D$ on $V_M$ fits
well to the exponential law $R_D \sim e^{\beta V_M}$, where $\beta
\simeq 14.2$ V$^{-1}$.
Since $d \simeq \alpha V_M$ ($\alpha < 0$), this is consistent with the
expected dependence $R_D \sim e^{-4 k_F d}$ which follows from the Bessel
function asymptotics at large arguments. Using the experimentally found
$\alpha$, we have $k_F \simeq 6.1 \times 10^4$ cm$^{-1}$ at the peak;
surprisingly this
corresponds to a low density of about only 8 electrons per 2 $\mu$m
wire segment! In the region of less negative voltages, $V_M > -0.7$ V, we see a
decrease of $R_D$. This occurs in the regime in which the tunneling leads to
penetration of a considerable fraction of the current from the drive wire to the
drag one and thus reduces the measured $R_D$.

Figure 4 shows $V_D$ vs $V_{T}$ for $V_{DS}=$ 300 $\mu$V, $B=0$, and different
temperatures. We see again the general pattern of Fig. 2 and a
decrease in the drag response with increasing temperature. The corresponding
decrease of the peak value of $R_D$ with $T$ is shown in the inset.
For $0.2 \leq T \leq 1$ K the temperature dependence can be described well
by the {\it power law} $R_D \propto T^{x}$ with $x=-0.77(2)$. This behavior is
in sharp contrast with the {\it linear} behavior (x=+1.0) expected
[9]-\cite{10a} for Fermi liquids. As we move to the right shoulder of the peak
at less negative $V_T$, the temperature dependence of the drag signal becomes
progressively weaker and is again described well by a power law. For example, at
$V_T=-1.17$ V we have found $x=-0.61(2)$. The presence of the magnetic field
up to 0.86 T does not change the observed behavior: at 0.86 T we
obtained $x=-0.73(6)$ at the peak.

The unusual temperature dependence cannot be attributed to a
temperature-induced modification of the wire conductance, since
the latter is found to be almost unchanged in this range of temperatures.
Possible reduction of the interwire Coulomb coupling due to enhanced
screening by the reservoirs and gates seem to be unlikely at such small
temperatures. On the other hand, one can argue that correlated liquid
behavior is established in the wires. Indeed, it is hardly
surprising that the temperature dependence of the observed CD
does not fit into a FL scenario, because for the peak conditions
the ratio of the mean distance between the electrons within one wire
to the Bohr radius $r_s=\bar{r}/a_B \simeq 26$ is large. Below we find
that the temperature dependence of $R_D$ 
is in good agreement with a theory of CD between TL liquids.

The smallness of the drag resistance ($R_D < 100~\Omega$) indicates
a weak interwire backscattering coupling. In this case $R_D$ should obey a
power law as long as the thermal length $L_T$ is well in between the wire
length $L=$2 $\mu$m and the mean electron distance $\bar r \simeq 250$ nm.
The exponent $x$ is determined by the TL parameter $K_{c-}$
of the relative charge mode [7]. For spin-unpolarized electrons,
as in the present experiment, it is
\begin{equation}\label{exponent}
x = 2 K_{c-} -1.
\end{equation}
As $L_T$ approaches $L$, the temperature dependence of $R_D$ is
expected to weaken.

Let us first see whether the condition $\bar r < L_T < L$ is fulfilled in
our experiment. If Eq. (1) holds, the parameter $K_{c-}$
can be determined from the experimental data as
$K_{c-}=0.12-0.2$ depending on $V_T$.
Given a Fermi wavevector $k_F \simeq 6~\mu$m$^{-1}$, we find that
$L_T = \hbar v_F/K_{c-} k_B T=  
L=$2 $\mu$m at a temperature of $\simeq 250$ mK, and that $L_T$ approaches
$\bar r \simeq 250$ nm for temperatures of order 2 K. This means that
there is a narrow temperature window in which a power-law behavior
of $R_D(T)$ might be expected. The data is indeed consistent with such a
power-law dependence of $R_D(T)$ for temperatures in the range
$0.2-1$ K, while at lower temperatures a weakening of the drag is
observable, see inset of Fig. 4. It is crucial, however, to check
if the rather low values of $K_{c-}$ obtained are consistent with
the system parameters, which we do next.

It was shown recently [16] that
the interaction parameter of a single quantum wire, calculated
by standard perturbative methods, yields reliable values even
for small values of $k_F w$ ($w$ is the width of the wire) down to
0.1, while in our experiment $w \simeq 23$ nm (determined from experimental
data) and $k_F w \simeq 0.14$. Encouraged by this result, we
determine $K_{c-}$ in a similar way via the compressibility of the
relative charge mode obtained in the Hartree-Fock approximation; this leads
to
\begin{equation}
K_{c-}^{HF} = \left(1 + [2( V_0 - {\bar V}_0) -V_{2 k_F}]/\pi v_F
\right)^{-1/2},
\end{equation}
where $V$ and $\bar V$ denote intra- and inter-wire potentials,
respectively. Modeling the potentials by $V =e^2\epsilon^{-1}(x^2+w^2)^{-1/2}$
and $\bar V=e^2\epsilon^{-1}(x^2+d^2)^{-1/2}$,
we obtain an interaction parameter of $K_{c-}^{HF}=0.178$.
To obtain this result we used the parameters
$d=200$ nm,
determined experimentally as
well as estimated from electrostatic calculations of the double-well potential
profile created by the three parallel infinite gates for $V_M=-0.8$ V
and $V_T=V_B=-1.5$ V, $w=23$ nm, $k_F = 6.1$ $\mu$m$^{-1}$, $\epsilon
= 12.5$, and
$m^* = 0.068~m_e$. If we take into account screening by a homogeneous gate,
i.e., if we subtract the image-charge potentials
$e^2\epsilon^{-1}(x^2+w^2+4l^2)^{-1/2}$
and $e^2\epsilon^{-1}(x^2+d^2+4l^2)^{-1/2}$ ($l=80$ nm), from 
$V$ and $\bar{V}$, respectively, we obtain $K_{c-}^{HF(S)}$ = 0.212. Since
in reality we have split gates, the true value of $K_{c-}$ is expected to be
somewhere between $K_{c-}^{HF}$ and $K_{c-}^{HF(S)}$ and is in reasonable
agreement with the experimental value. The drag resistance itself is
proportional [7] to the square $|\bar V_{2k_F}|^2$ of the $2k_F$-component
of the interwire interaction, which leads to the exponential dependence
$R_D \sim e^{-4 k_F d}$ for 
$d > k_F^{-1}$, 
whether the screening is present or not. Thus, we stress that our previous
analysis based on this dependence remains valid in the TL liquid approach.

The negative power-law temperature dependence is not the only signature
against a FL drag theory we obtained in our experiment. The experimental
peak value
of $R_D$, at $T=$60 mK,  is more than one order of magnitude larger
than the value obtained using equations of the FL theory for the ballistic
transport regime [10], \cite{10a}. That the measured drag is larger could be
explained by the interaction renormalized interwire backscattering
probability, which should be larger than the bare one.

Considering the large interwire separation in our experiment, one cannot
rule out a possibility of a phonon-mediated drag (PMD) contribution to $R_D$.
Existing results for 2D systems [17,8] show that the PMD rapidly decreases with
$T$ at $T < 2$ K and depends rather weakly on the interlayer separation. Since
our data qualitatively contradict such a behavior, we conclude that the PMD,
if any, does not play a major role in our measurements. Since little
information is available on PMD in 1D systems, further discussion of this
subject can only be of purely speculative nature and is not appropriate here.

In conclusion, we have investigated the Coulomb drag between 1D electron systems
and  observed a {\it negative power-law temperature dependence of the drag
resistance}, which can be explained quantitatively in terms of the TL
liquid concept. Clearly, further 
work is necessary to put the TL nature of the CD on a firm footing.

We are grateful to A. Stern and V. L. Gurevich for fruitful discussions.
P. V. acknowledges support from the NSERC Grant No. OGP0121756.



\begin{figure}

\caption{Schematics of the device. The letters T, M, and B denote
upper, middle and lower gates.}
\label{fig.1}

\ \\

\caption{Drag voltage $V_D$ and drive wire conductance $G$
vs voltage of the upper gate $V_T$ at $T=70$ mK, $V_M=-0.74$ V,
$V_B=-1.525$ V, and
$V_{DS}=300~\mu$V. The dot-dashed and dashed lines show the
staircases $G(V_T)$ in
magnetic fields of 0.35 T and 0.86 T, respectively. The inset shows
the dependence
of the first peak in $V_{D}$ on $V_{DS}$. }
\label{fig.2}

\ \\

\caption{(a) Drag voltage $V_D$ vs voltage of the middle
gate for $T=60$ mK. (b) The logarithm of $R_D$ vs $V_M$.
The dotted curve shows the exponential decay of $R_D$ with $V_M$. }
\label{fig.3}

\ \\

\caption{The same as in Fig. 2 for temperatures 70, 180, 300, and 600 mK,
corresponding to the curves in a descending order. The inset shows the
peak drag resistance $R_D$ vs temperature $T$ and the dotted curve
is a power-law fit. }
\label{fig.4}

\ \\

\end{figure}


\begin{references}

\bibitem[\dagger]{vol} Permanent address: Institute of Solid State Physics RAS,
Chernogolovka, Moscow region, 142432, Russia.
\bibitem[\ddagger]{ol} Permanent address: Institute of Semiconductor Physics,
NAS Ukraine, Prospekt Nauki 45, Kiev, 03028, Ukraine.


\bibitem{1} J. Voit. Rep. Progr. Phys. {\bf 57}, 977 (1994).

\bibitem{2} W. Apel and T. M. Rice, Phys. Rev. B {\bf 26}, 7063 (1982).

\bibitem{3} S. Tarucha {\it et al.}, 
Sol. State. Commun. {\bf 94}, 413 (1995); A. Yacoby {\it et al.},
Phys. Rev. Lett. {\bf 77}, 4612 (1996).

\bibitem{4} I. Safi and H.J. Schulz, Phys. Rev. B {\bf 52}, R17040 (1995);
V. V. Ponomarenko, ibid 
{\bf 52}, R8666 (1995); Yu. Oreg and A. M. Finkel'stein, ibid 
{\bf 54}, R14265 (1996). 

\bibitem{5} K. Flensberg, Phys. Rev. Lett. {\bf 81}, 184 (1998).

\bibitem{6} Yu. A. Nazarov and D. V. Averin, Phys. Rev. Lett.
{\bf 81}, 653 (1998).

\bibitem{7} R. Klesse and A. Stern, Phys. Rev. B
{\bf 62}, 16912 (2000).

\bibitem{8} 
A. Rojo, J. Phys. Cond. Matt. {\bf 11}, R31 (1999).

\bibitem{9}  Yu. M. Sirenko  and P. Vasilopoulos, Phys. Rev. B
{\bf 46}, 1611 (1992). 
\bibitem{10} V. L. Gurevich  {\it et al.},  
J. Phys.: Cond. Matter {\bf 10}, 2551 (1998).
\bibitem{10a} O.  E.  Raichev and
P. Vasilopoulos, Phys. Rev. B {\bf 61}, 7511 (2000).

\bibitem{11} C. L. Kane and M. P. A. Fisher, Phys. Rev. Lett. {\bf 68},
1220 (1992). 

\bibitem{12} V. V. Ponomarenko and D. V. Averin, Phys. Rev. Lett.,
{\bf 85}, 4928 (2000).

\bibitem{13} P. Debray {\it et al.},
Physica E {\bf 6}, 694 (2000).

\bibitem{14} L.P. Kouwenhoven {\it et al.},
Phys. Rev. B {\bf 39}, 8040 (1989).

\bibitem{15} C. E. Greffield {\it et al.}, 
cond-mat/0005354.


\bibitem{16} H. C. Tso {\it et al.}, 
Phys. Rev. Lett. {\bf 68}, 2516 (1992); \\T. G. Gramila
{\it et al.},
Phys. Rev. B {\bf 47}, 12957 (1993).

\end{references}
\end{document}